# Improved Estimation of the Spectral Efficiency Versus Energy-Per-Bit Tradeoff in the Wideband Regime

Richard J. Barton
NASA Johnson Space Center, Houston, TX

*Abstract*—In this paper, a new lower bound on spectral efficiency in the low-power wideband regime is derived and utilized to develop an improved estimate of the behavior of spectral efficiency above but in the vicinity of the Shannon limit.

*Index Terms*—Wideband Spectral Efficiency

## I. INTRODUCTION

In a comprehensive study published in 2002, Verdu [1] investigated the tradeoff between spectral efficiency (b/s/Hz) and transmitted energy per information bit in the wideband, power-limited regime. In that work, the fundamental bandwidth-power tradeoff was characterized by approximating spectral efficiency as an affine function of normalized energy-per-bit $E_b/N_0$ (measured in dB), in which the two quantities of interest are the minimum possible value of $E_b/N_0$ required for reliable communication (i.e., the Shannon limit, $E_b/N_{0\min}$) and the wideband slope of the spectral efficiency curve at $E_b/N_{0\min}$ (measured in b/s/Hz/dB). It was also demonstrated that both the Shannon limit and the wideband slope are determined by the first two derivatives of Shannon's capacity function $C(\text{SNR})$ evaluated at $\text{SNR} = 0$. Based on this, the notion of second-order (or wideband) optimality of an input distribution was defined as the property that the mutual information between the channel input and output (as a function of SNR) with respect to the given input distribution achieves the same first two derivatives as the capacity function at the value $\text{SNR} = 0$. Hence, the second-order optimality of an input distribution implies that the given input distribution will achieve the same bandwidth-power tradeoff in the wideband regime as the capacity-achieving input distribution. Finally, utilizing the notion of second-order optimality and wideband slope, [1] was able to characterize the bandwidth-power tradeoff and the wideband-optimal distributions for several important coherent and non-coherent (fading) Gaussian noise channels.

While the wideband approximation for spectral efficiency developed in [1] can certainly be used in a practical context to approximate the attainable spectral efficiency for large values of bandwidth that are nevertheless finite, a couple of questions remain. For one, it is of interest to know whether the affine approximation based on the Shannon limit and wideband slope is a true upper bound for the spectral efficiency function in a vicinity of the Shannon limit, as it would be if the spectral efficiency function were concave in some neighborhood of the Shannon limit. Similarly, it is of interest to know whether the approximation can be improved for finite values of bandwidth by incorporating additional derivative information from the capacity function and extending the Taylor series expansion of the spectral efficiency function beyond first order. In this paper, we show that the answer to both of these questions is no. That is, we show that the spectral efficiency function has no derivatives higher than first order at $E_b/N_{0\min}$ nor is it concave in any neighborhood of $E_b/N_{0\min}$. Nevertheless, we also show that it is possible to develop a new approximation of spectral efficiency in the wideband regime, which is not affine but is derived from the value of $C(\text{SNR})$ and its first two derivatives at $\text{SNR} = 0$ and is asymptotically equivalent to both the true spectral efficiency and the affine approximation given in [1] at $E_b/N_{0\min}$. We show that this new approximation provides a lower bound to the true spectral efficiency in some neighborhood of $E_b/N_{0\min}$ where the previous affine approximation provides an approximate upper bound. Hence, the two approximations can be used to bound the true spectral efficiency both above and below in some neighborhood of $E_b/N_{0\min}$, and they can be averaged to provide a third estimate that is generally superior to both in the same neighborhood. Although this does not imply that it is possible to estimate spectral efficiency accurately for arbitrary channels in regions far from the Shannon limit, it does lead to an approach that works well for the simple additive white Gaussian noise (AWGN) channel and provides a tool that may prove more useful than just the wideband slope for analyzing the bandwidth-power tradeoffs on many different channels for achievable (i.e., finite) values of bandwidth.

## II. DEFINITIONS, NOTATION, AND ASSUMPTIONS

Throughout this paper, we closely follow the notation adopted in [1]. We adopt as our reference model the simple

complex-valued, baseband AWGN model given by
$$y(t) = Ax(t) + n(t),$$
where $A \neq 0$ represents a fixed channel gain, which is assumed known at both transmitter and receiver, and $n(t)$ is a proper complex-valued AWGN process with power spectral density $N_0$, as defined in [2]. We assume that the transmitted signal $x(t)$ has two-sided bandwidth $0 < B < \infty$ and is sampled at the Nyquist rate of $B$ samples per second to produce the equivalent discrete-time observational model
$$y_k = Ax_k + n_k,$$
where the discrete-time AWGN process now has variance $\sigma^2 = BN_0$.

In this case, the Shannon capacity for the channel is given by the well known formula
$$C(\xi) = \log_2 e \cdot \ln(1 + A\xi),$$
where $C(\xi)$ represents the maximum achievable rate on the channel, measured in bits per symbol for the discrete-time channel and in bits per second per Hertz (b/s/Hz) for the continuous time channel, and $\xi$ represents the signal-to-noise ratio (SNR) on the channel given by $\xi = P/BN_0$, where $P$ is the average transmitted signal power on the channel. Note that $C(\xi)$ is equivalent to the spectral efficiency of the channel.

Another quantity of interest is $C(\xi)/\xi$, which represents the average achievable rate on the channel per unit of transmitted SNR. It can be shown [3] that the quantity
$$\lim_{\xi \to 0} \frac{C(\xi)}{\xi} = C'(0)$$
represents the capacity per unit cost on the channel, which is the maximum achievable data rate on the channel per unit of SNR. Conversely, the function
$$\frac{E_b}{N_0} = h(\xi) = \frac{\xi}{C(\xi)}$$
represents the average normalized energy required to reliably transmit a bit for a particular value of SNR, and it follows that the minimum value of $E_b/N_0$ required to transmit a single bit of information reliably on the channel is given by
$$\frac{E_b}{N_0}_{\min} = \lim_{\xi \to 0} h(\xi) = \frac{1}{C'(0)}.$$

Note that these definitions imply that $h(\xi)C(\xi) = \xi$. Note also that the properties of Shannon capacity, or more generally the capacity cost function [4], guarantee that for any physically realizable channel, we have $C(0) = 0$ and $0 \leq C'(0) < \infty$. Throughout the rest of this paper, we make the additional simplifying regularity assumptions that $C(\xi)$ is twice continuously differentiable on some interval $0 \leq \xi \leq \xi_0$ and that $C'(0) > 0$ and $-\infty < C''(0) < 0$. Finally, when referring to functions of $E_b/N_0$ measured in dB, we will let $\gamma_b = 10\log_{10}(E_b/N_0)$ and $\gamma_{b,\min} = 10\log_{10}(E_b/N_{0\min})$.

### III. RESULTS

The region of interest in this paper, called the *wideband regime*, is the operating region in which the available bandwidth is very large relative to the normalized transmitted power, the spectral efficiency is close to zero, and the $E_b/N_0$ required for reliable communication is close to $E_b/N_{0\min}$. By definition, this region is power constrained rather than bandwidth constrained, but bandwidth still has a cost, and it is of interest to understand the relationship between bandwidth, achievable data rate, and SNR for finite values of the bandwidth in the wideband regime. This relationship is captured in the functional dependence of spectral efficiency on $E_b/N_0$, and the goal of this paper is to derive improved approximations and bounds for this functional dependence in the wideband regime.

Under our assumptions regarding the function $C(\xi)$, it can be shown that both $h(\xi)$ and $C(\xi)$ are continuously invertible in the vicinity of $\xi = 0$. It follows that we can solve for $C$ as a function of $h$ in the vicinity of the point $h(0) = 1/C'(0) = E_b/N_{0\min}$. We denote this function by $\mathsf{C}(E_b/N_0)$ and its equivalent function of $\gamma_b$ as $\tilde{\mathsf{C}}(\gamma_b)$, and we seek good approximations for $\mathsf{C}(E_b/N_0)$ and $\tilde{\mathsf{C}}(\gamma_b)$ in the vicinity of $E_b/N_{0\min}$. Toward this end, it has been shown in [1] that the derivative of $\mathsf{C}(E_b/N_0)$ at the point $E_b/N_{0\min}$, which Verdu refers to as the *wideband slope* of the channel, is given by
$$\mathsf{C}'\left(\frac{E_b}{N_{0\min}}\right) = \frac{-2[C'(0)]^3}{C''(0)}, \quad \tilde{\mathsf{C}}'(\gamma_{b,\min}) = \frac{-\ln 10 [C'(0)]^2}{5C''(0)}.$$

Hence, one approach to approximating $\mathsf{C}(E_b/N_0)$ in the wideband regime is to use the affine approximation for $\tilde{\mathsf{C}}(\gamma_b)$, which results in
$$\tilde{\mathsf{C}}(\gamma_b) \approx \tilde{\mathsf{C}}_1(\gamma_b) = \frac{-\ln 10 [C'(0)]^2}{5C''(0)}(\gamma_b - \gamma_{b,\min}), \forall \gamma_b \geq \gamma_{b,\min},$$
$$\mathsf{C}\left(\frac{E_b}{N_0}\right) \approx \mathsf{C}_1\left(\frac{E_b}{N_0}\right) = \frac{-2\ln 10 [C'(0)]^2}{C''(0)}$$
$$\cdot \left(\log_{10}\frac{E_b}{N_0} - \log_{10}\frac{E_b}{N_{0\min}}\right), \forall \frac{E_b}{N_0} \geq \frac{E_b}{N_{0\min}}.$$

This is the approach taken in [1]. To derive a (hopefully) better approximation, we note that

$$h'(\xi) = \frac{C(\xi) - \xi C'(\xi)}{C^2(\xi)},$$

from which it follows that

$$h'(0) = \lim_{\xi \to 0} \frac{C(\xi) - \xi C'(\xi)}{C^2(\xi)} = \lim_{\xi \to 0} \frac{-\xi C''(\xi)}{2C(\xi)C'(\xi)} = \frac{-C''(0)}{2[C'(0)]^2}.$$

Hence, for values of $\xi \approx 0$, we can write

$$h(\xi) \approx h(0) + \xi h'(0) = \frac{1}{C'(0)} - \xi \frac{C''(0)}{2[C'(0)]^2}.$$

Using this relationship and the fact that $h(\xi)C(\xi) = \xi$, we then get

$$h(\xi) \approx \frac{1}{C'(0)} - h(\xi)C(\xi)\frac{C''(0)}{2[C'(0)]^2},$$

or

$$h \approx \frac{1}{C'(0)} - hC\frac{C''(0)}{2[C'(0)]^2},$$

from which we can solve for

$$C \approx \frac{2C'(0)}{C''(0)}\left(\frac{1}{h} - C'(0)\right).$$

This gives the approximation

$$\mathsf{C}\left(\frac{E_b}{N_0}\right) \approx \mathsf{C}_2\left(\frac{E_b}{N_0}\right) = \frac{2C'(0)}{C''(0)}\left[\left(\frac{E_b}{N_0}\right)^{-1} - C'(0)\right], \forall \frac{E_b}{N_0} \geq \frac{E_b}{N_{0\min}}.$$

$$\tilde{\mathsf{C}}(\gamma_b) \approx \tilde{\mathsf{C}}_2(\gamma_b) = \frac{2C'(0)}{C''(0)}\left[10^{-\gamma_b/10} - C'(0)\right], \forall \gamma_b \geq \gamma_{b,\min}.$$

Before comparing the behavior of these approximations, we note the following simple result.

*Lemma.* Let $h = h(\xi) = E_b/N_0$ and $h_{\min} = h(0) = E_b/N_{0\min}$. The functions $\mathsf{C}(h)$ and $\mathsf{C}_2(h)$ must have the same first derivative at $h = h_{\min}$; that is, $\mathsf{C}'(h_{\min}) = \mathsf{C}'_2(h_{\min})$. Furthermore, we must have $\mathsf{C}(h) \geq \mathsf{C}_2(h)$ in a neighborhood of $h = h_{\min}$.

*Proof.* Note that the approximation $h(\xi) \approx h(0) + \xi h'(0)$ can be written more precisely as

$$h(\xi) = h(0) + \xi h'(0) + \xi O(\xi),$$

where the term $O(\xi)$ represents an unknown function of $\xi$ that satisfies $\lim_{\xi \to 0} O(\xi) = 0$. This implies that the approximation $\mathsf{C}(h) \approx \mathsf{C}_2(h)$ can then also be written more precisely as

$$\mathsf{C}(h) = \frac{\mathsf{C}_2(h)}{1 + O(\xi)} = \frac{2C'(0)}{C''(0)}\left[\frac{1}{h} - C'(0)\right] \cdot \frac{1}{1 + O(\xi)}.$$

Now, since the function $h(\xi)$ is continuously invertible in a neighborhood of $\xi = 0$, we have (abusing notation slightly) $\xi(h) = h^{-1}(h)$, and $\lim_{h \to h_{\min}} \xi(h) = 0$. It follows that

$$\mathsf{C}'(h_{\min}) = \lim_{\varepsilon \to 0} \frac{\mathsf{C}(h_{\min} + \varepsilon)}{\varepsilon} = \lim_{\varepsilon \to 0} \frac{\mathsf{C}_2(h_{\min} + \varepsilon)}{\varepsilon[1 + O(\xi(h_{\min} + \varepsilon))]}$$

$$= \lim_{\varepsilon \to 0} \frac{\mathsf{C}_2(h_{\min} + \varepsilon)}{\varepsilon} = \mathsf{C}'_2(h_{\min}).$$

This proves the first assertion. To prove the second, we note that

$$h''(\xi) = \frac{-\xi C(\xi)C''(\xi) - 2C(\xi)C'(\xi) + 2\xi[C'(\xi)]^2}{C^3(\xi)},$$

and

$$\lim_{\xi \to 0} h''(\xi) = \lim_{\xi \to 0} \frac{-\xi C(\xi)C''(\xi) - 2C(\xi)C'(\xi) + 2\xi[C'(\xi)]^2}{C^3(\xi)}$$

$$= \lim_{\xi \to 0} \frac{\begin{pmatrix} -\xi C'(0)C''(0) - 2C'(0)[C'(0) + \xi C''(0)] \\ +2[C'(0) + \xi C''(0)]^2 \end{pmatrix}}{\xi^2[C'(0)]^3}$$

$$= \lim_{\xi \to 0} \frac{C''(0)}{\xi[C'(0)]^2} = -\infty.$$

Hence, it follows from Taylor's Theorem [5] that the term $O(\xi)$ must satisfy $O(\xi) < 0$ in some neighborhood of $h = h_{\min}$, and

$$\mathsf{C}(h) = \frac{\mathsf{C}_\infty(h)}{1 + O(\xi)} \geq \mathsf{C}_2(h)$$

in that neighborhood. This proves the second assertion. ∎

We can now derive directly that

$$\mathsf{C}'\left(\frac{E_b}{N_{0\min}}\right) = \mathsf{C}'_2\left(\frac{E_b}{N_{0\min}}\right) = \frac{-2[C'(0)]^3}{C''(0)},$$

$$\tilde{\mathsf{C}}'(\gamma_{b,\min}) = \tilde{\mathsf{C}}'_2(\gamma_{b,\min}) = \frac{-\ln 10[C'(0)]^2}{5C''(0)},$$

which is just a restatement of the wideband slope as derived in [1] but derived independently starting with the approximation $\mathsf{C}_2(h)$. Both estimates depend only on the first two derivatives of the capacity function $C(\xi)$ at $\xi = 0$, and they are asymptotically equivalent to each other and to $\mathsf{C}(E_b/N_0)$ in the vicinity of $E_b/N_0 = E_b/N_{0\min}$. However, they can behave quite differently for values of $E_b/N_0 > E_b/N_{0\min}$ (i.e., for values of $B < \infty$).

To see this, consider the simple AWGN channel with gain $A = 1$. In this case, we can find the true value of $\mathsf{C}(E_b/N_0)$ by solving the equation

$$\mathsf{C} = \log_2 e \cdot \ln\left(1 + \mathsf{C}\frac{E_b}{N_0}\right)$$

numerically in an arbitrary region $E_b/N_0 \geq E_b/N_{0\min}$. Fig. 1 illustrates the performance of $\mathsf{C}_1(E_b/N_0)$, $\mathsf{C}_2(E_b/N_0)$, and $\mathsf{C}(E_b/N_0)$ over the interval $E_b/N_{0\min} \leq E_b/N_0 \leq 10 E_b/N_0$ for the AWGN channel.

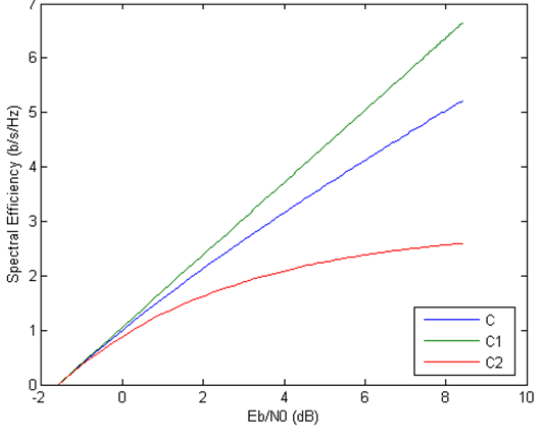

Fig. 1. Comparison of True Spectral Efficiency for the AWGN Channel with 2 Approximations

As the figure illustrates, both of the approximations perform very well for values of $E_b/N_0$ near $E_b/N_{0\min}$, but they diverge significantly from each other and the true spectral efficiency as $E_b/N_0$ grows. Nevertheless, $\mathsf{C}_1(E_b/N_0)$ and $\mathsf{C}_2(E_b/N_0)$ bracket the true spectral efficiency nicely to provide upper and lower bounds, respectively, for values of $E_b/N_0$ over the entire 10 dB range of values illustrated. Furthermore, if we define $\hat{\mathsf{C}}(E_b/N_0)$ as the average of $\mathsf{C}_1(E_b/N_0)$ and $\mathsf{C}_2(E_b/N_0)$, then we get an improved estimate of the true spectral efficiency over the entire range, as illustrated in Fig. 2.

## IV. DISCUSSION

It should be noted that although the average of $\mathsf{C}_1(E_b/N_0)$ and $\mathsf{C}_2(E_b/N_0)$ as derived above does provide a significantly better estimate for spectral efficiency than either $\mathsf{C}_1(E_b/N_0)$ or $\mathsf{C}_2(E_b/N_0)$ for the AWGN channel, that will not necessarily be the case in general. Furthermore, although it appears to be true for the AWGN channel, it is not actually the case that $\mathsf{C}_1(E_b/N_0)$ provides a true upper bound for $\mathsf{C}(E_b/N_0)$ in a neighborhood of $E_b/N_{0\min}$. In particular, it follows from the properties of the second derivative of $h(\xi)$ derived in the Lemma and the well known properties of derivatives of function compositions and inverses [6] that

$$\lim_{E_b/N_0 \to E_b/N_{0\min}} \mathsf{C}''(E_b/N_0) = \infty.$$

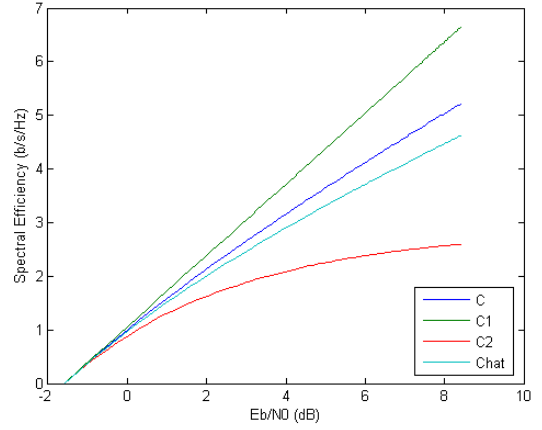

Fig 2. Comparison of True Spectral Efficiency for the AWGN Channel with Improved Approximation

Hence, $\mathsf{C}(E_b/N_0)$ is neither concave nor bounded above by $\mathsf{C}_1(E_b/N_0)$ in a neighborhood of the Shannon limit, nor does it have a meaningful power series expansion beyond $\mathsf{C}_1(E_b/N_0)$ around the Shannon limit. It follows that the nonlinear behavior of $\mathsf{C}(E_b/N_0)$ in the wideband regime cannot be accurately estimated solely from values of its derivatives at $E_b/N_{0\min}$. Having said that, recall that we have already shown that $\mathsf{C}_2(E_b/N_0) \leq \mathsf{C}(E_b/N_0)$ in some neighborhood of $E_b/N_{0\min}$, or equivalently, that $\tilde{\mathsf{C}}_2(\gamma_b) \leq \tilde{\mathsf{C}}(\gamma_b)$ in some neighborhood of $\gamma_{b,\min}$, as illustrated in Fig. 1. On the other hand, it follows as in the proof of the Lemma that

$$\tilde{\mathsf{C}}(\gamma_b) = \tilde{\mathsf{C}}_1(\gamma_b) + O(\xi(\gamma_b))(\gamma_b - \gamma_{b,\min})$$
$$= \left[\tilde{\mathsf{C}}_1'(\gamma_b) + O(\xi(\gamma_b))\right](\gamma_b - \gamma_{b,\min}),$$

where $\lim_{\gamma_b \to \gamma_{b,\min}} O(\xi(\gamma_b)) = 0$. Hence, for any $\varepsilon > 0$, there is a neighborhood of $\gamma_{b,\min}$ for which $|O(\xi(\gamma_b))| < \varepsilon$, and it follows that

$$\tilde{\mathsf{C}}(\gamma_b) \leq \left[\tilde{\mathsf{C}}_1'(\gamma_b) + \varepsilon\right](\gamma_b - \gamma_{b,\min}),$$

for all $\gamma_{b,\min}$ in that neighborhood. If we let $\tilde{\mathsf{C}}_{1,\varepsilon}(\gamma_b)$ represent the line $\tilde{\mathsf{C}}_1(\gamma_b)$ with slope increased by $\varepsilon$, then for each such $\varepsilon$, we can always find a neighborhood such that

$$\tilde{\mathsf{C}}_2(\gamma_b) \leq \tilde{\mathsf{C}}(\gamma_b) \leq \tilde{\mathsf{C}}_{1,\varepsilon}(\gamma_b).$$

Hence, for reasonably well behaved functions $\tilde{\mathsf{C}}(\gamma_b)$, it is quite plausible that the approximation $\tilde{\mathsf{C}}_1(\gamma_b)$ will provide an approximate affine upper bound for $\tilde{\mathsf{C}}(\gamma_b)$ over a large subset of the region $\gamma_b \geq \gamma_{b,\min}$, and that $\tilde{\mathsf{C}}_2(\gamma_b)$ will provide a nonlinear lower bound over the same region. Hence, the pair $\tilde{c}_1(\gamma_b)$ and $\tilde{c}_2(\gamma_b)$ will bracket the behavior of $\tilde{c}(\gamma_b)$ over the region, and they can (perhaps) be averaged to provide an improved estimate of $\tilde{c}(\gamma_b)$.

## V. Conclusion

In this correspondence, we have investigated possible improvements to estimation of the spectral efficiency as a function of energy-per-bit in the wideband regime. It was first established in [1] that the asymptotic behavior of spectral efficiency in the vicinity of the Shannon limit $E_b/N_{0\min}$ is completely characterized by the so-called wideband slope, i.e., the slope of spectral efficiency evaluated at $E_b/N_{0\min}$, which is in turn determined by the first two derivatives of the capacity function $C(\text{SNR})$ at $\text{SNR} = 0$. We have shown in this work that further characterization of the nonlinear behavior of spectral efficiency in the wideband regime, i.e., beyond the linear behavior captured by the wideband slope, is not really feasible based on knowledge of the higher-order derivatives of spectral efficiency at $E_b/N_{0\min}$. Nevertheless, as we have also shown, it is possible to improve the affine approximation of spectral efficiency defined by the wideband slope by deriving a new approximation for spectral efficiency in the wideband regime that is guaranteed to be a lower bound for the true spectral efficiency within some neighborhood of $E_b/N_{0\min}$. This lower bound, along with an upper bound derived by increasing the slope of the original affine estimate incrementally, can then be used to bracket the true value of spectral efficiency in the same neighborhood of $E_b/N_{0\min}$, and an improved estimate of spectral efficiency can be derived simply by averaging the upper and lower bounds. This behavior has been illustrated using the AWGN channel as an example.

Although the results presented here do not imply that it is possible to estimate spectral efficiency accurately for arbitrary channels in regions far from the Shannon limit, it does lead to an approach that works well for the simple AWGN channel and can be expected to work well for variations of that channel as well, such as Gaussian fading channels. It is hoped that these results will provide an additional tool that may prove more useful than the wideband slope alone for analyzing the bandwidth-power tradeoffs on many different channels for achievable (i.e., finite) values of bandwidth.